\documentclass[a4paper]{article}
\usepackage[a4paper,top=3cm,left=2cm,right=2cm,bottom=3cm]{geometry}
\usepackage{amsmath}
\usepackage{amsthm}
\usepackage[latin1]{inputenc}
\usepackage{graphicx}
\usepackage{natbib}
\usepackage{algorithm}
\usepackage{setspace}
\usepackage{textpos}
\usepackage{fancyhdr}
\usepackage{graphicx}
\usepackage{subfig}
\usepackage{rotating}
\usepackage{lscape}
\usepackage{amsmath}
\usepackage{amsthm}
\usepackage{algorithm}
\usepackage[noend]{algpseudocode}
\usepackage{setspace}
\usepackage{graphicx,color}
\usepackage{hyperref}
\usepackage{enumitem}
\usepackage{amssymb}

\newtheorem{thm}{Theorem}
\newenvironment{myproof}
    {\noindent {\textsc{Proof}. \hspace{0pt}}}
    {\vspace{5pt}}

\newcommand{\up}[1]{\raisebox{1.3ex}[0pt]{#1}}

\begin{document}
\large

\title{A speed and departure time optimization algorithm for the Pollution-Routing Problem}

\author{\textbf{Raphael Kramer} \\
Dipartimento di Scienze e Metodi dell'Ingegneria \\ Università degli Studi di Modena e Reggio Emilia, Italy \\ 
raphael.kramer@unimore.it
\and
\textbf{Nelson Maculan} \\
Departamento de Engenharia de Sistemas e Computação \\ Universidade Federal do Rio de Janeiro, Brazil \\
maculan@cos.ufrj.br
\and
\textbf{Anand Subramanian} \\
Departamento de Engenharia de Produção \\ Universidade Federal da Paraíba, Brazil \\
anand@ct.ufpb.br
\and
\textbf{Thibaut Vidal} \\
Departamento de Informática \\ Pontifícia Universidade Católica do Rio de Janeiro, Brazil \\
vidalt@inf.puc-rio.br
}
\date{}

\maketitle

\begin{textblock*}{\textwidth}(0cm,-14cm)
\noindent ©2018. This manuscript version is made available under the CC-BY-NC-ND 4.0 license http://creativecommons.org/licenses/by-nc-nd/4.0/.
\end{textblock*}

\vspace{-0.25cm}
\begin{center}
Author accepted manuscript (AAM), published in \\
``\emph{European Journal of Operational Research}, 2015, 247(3):782--787''.\\ 
DOI: \url{https://doi.org/10.1016/j.ejor.2015.06.037}.\\ \vspace{0.1cm}
\end{center}
\vspace{0.675cm}

\begin{abstract}

We propose a new speed and departure time optimization algorithm for the Pollution-Routing Problem (PRP), which runs in quadratic time and returns a certified optimal schedule. 
This algorithm is embedded into an iterated local search-based metaheuristic to achieve a combined speed, scheduling and routing optimization. 
The start of the working day is set as a decision variable for individual routes, thus enabling a better assignment of human resources to required demands.
Some routes that were evaluated as unprofitable
can now appear as viable candidates later in the day, leading to a larger search space and further opportunities of distance optimization via better service consolidation.
Extensive computational experiments on available PRP benchmark instances demonstrate the good performance of the algorithms. The flexible departure times from the depot contribute to reduce the operational costs by $8.36\%$ on the~considered~instances.

\end{abstract}

\onehalfspace

\section{Introduction} 
The Pollution-Routing Problem (PRP) is a variant of the Vehicle Routing Problem with environmental considerations, introduced in \cite{BektasLaporte2011}, and aiming to minimize operational and environmental costs subject to vehicle capacity and hard time-window constraints.
The costs are based on driver wages and fuel consumption, evaluated as a non-linear function of the distance traveled, vehicle load, and vehicle speed.
Some recent articles have proposed heuristics for the PRP: an adaptive large neighborhood search in \citet{DemirBL12}, and an ILS with a set-partitioning matheuristic in \citet{KramerSVC14}. Other contributions \citep{BektasLaporte2011,Franceschettietal2013,Dabiaetal2014}  focused on mathematical formulations and integer programming algorithms based on branch-and-price.

Vehicle-speed decisions play an important role in the PRP, since they do not only affect the total cost, but also the travel times between the locations, with a large impact on time-window feasibility. For this reason, most algorithms for the PRP perform -- at regular times during the search -- an optimization of vehicle speeds for the current routes. The resulting speed optimization subproblem (SOP), seeks to find the most cost-efficient arc speeds on a given route while respecting arrival-time constraints at each customer.

Some algorithms for the SOP have been recently proposed \citep{Norstadetal2011, DemirBL12,  KramerSVC14, Hvattumetal2013}. These algorithms run in quadratic time, consider identical cost/speed functions for each arc, and assume that the departure time is fixed.
Now, considering that the start of the working day (the departure time from the depot) is a decision variable leads to different optimality conditions and speed decisions. This may open the way to significant cost reductions, but also increases resolution complexity. Fewer articles have addressed this aspect.
In \citet{Dabiaetal2014}, the departure time from the first customer is optimized by means of a golden section search, within a pricing algorithm.
\cite{Franceschettietal2013} address the PRP with time-dependent travel times, a generalization of the considered problem. The resulting speed optimization algorithm is, however, more complex due to the presence of three time intervals with different speed/cost functions, involving 24 rules for speed choices. 
The solution is also not guaranteed to be optimal.
Finally, \citet{VidalJM14} showed that the optimal  SOP solution with deadlines and arc-dependent speed/cost functions can be achieved by solving a hierarchy of resource allocation problems.

This article contributes to the resolution of difficult vehicle routing variants with speed and departure time optimization.  We address the case where driver wages are computed from the departure time, hence allowing to better assign human resources to needed deliveries.
Note that a fixed departure time can still be obtained by reducing the departure time window to a point. 
Some routes that were evaluated as unprofitable with a fixed departure time policy can now appear as viable candidates, leading to a larger search space and further opportunities of routing optimization via better service~clustering.

We introduce a simple polynomial algorithm for the speed and departure time optimization (Section~\ref{Algorithm}), which runs in quadratic time. Moreover, we demonstrate the optimality of this algorithm (Section \ref{ProofSDTOA}).
The speed optimization is embedded into a vehicle routing matheuristic to produce high-quality routing plans. We conduct computational experiments on the classic PRP instances to evaluate the performance of the method, and assess the impact of departure time optimization on cost and pollution emissions (Section \ref{compExpSDTOA}). The results highlight very significant routing cost reduction, of $8.36\%$ on average. 
The CPU time of the new metaheuristic remains comparable to current state-of-the-art methods despite the fact that it deals with a more general problem.

\section{Problem Description}
\label{ProblemDescription}
The PRP with flexible departure times can be defined as follows. Let $G=(\mathcal{V},\mathcal{A})$ be a complete and directed graph with a set $\mathcal{V}=\{0,1,2,\dots,n\}$ of vertices and a set $\mathcal{A}=\{(i,j):i,j \in V, i \neq j\}$ of arcs. The vertex $\{0\}$ represents the depot, and the others are customer visits.
A set of $m$ vehicles with capacity $Q$ is available to service the customers.
Each customer $i$ has a non-negative demand~$q_i$, specified time-window interval $[a_i, b_i]$ to be serviced and a service time $\tau_{i}$. By convention, $q_0 = \tau_{0} = 0$ for the depot.
Each arc $(i,j) \in \mathcal{A}$ represents a travel possibility from node $i$ to $j$ for a distance $d_{ij}$, which can be traveled with any speed $v_{ij}$ in the interval $[v_{min},v_{max}]$.
The PRP aims at determining a speed matrix $(\mathbf{v})_{ij}$ for the arcs and a set of feasible routes $\mathbf{R}$ to serve all customers while minimizing environmental and operational costs.

Let $\boldsymbol\sigma = (\sigma_1,\sigma_2,\dots,\sigma_{|\sigma|})$ be a route, $f_{\sigma_i \sigma_{i+1}}$ be the vehicle load on arc $(\sigma_i,\sigma_{i+1})$ and~$t_{\sigma_i}$ be the arrival time at customer $\sigma_i$. The environmental cost is proportional to the fuel consumption, computed as in Equation (\ref{fc1}) where $w_1,w_2,w_3, w_4$ are parameters based on fuel properties, vehicle and network characteristics. The labor costs are proportional to route duration, computed as the difference between departure and arrival time $t_{\sigma_{|\sigma|}} - t_{\sigma_1}$.
Defining $\omega_{\textsc{fc}}$ as the fuel cost per liter and $\omega_{\textsc{dc}}$ as the driving cost per second, the objective of the PRP is given in Equation (\ref{PRPObj2}).
\begin{align}
F^\textsc{f}_{\sigma_{i}\sigma_{i+1}}(v_{\sigma_{i}\sigma_{i+1}}) &= d_{\sigma_{i}\sigma_{i+1}}  \left( \frac{w_1}{v_{\sigma_{i}\sigma_{i+1}}} + w_2 + w_3 f_{\sigma_{i}\sigma_{i+1}} + w_4v_{\sigma_{i}\sigma_{i+1}}^{2} \right) \label{fc1}
\end{align}
\begin{align}
Z_{\textsc{prp}}(\mathbf{R},\mathbf{v}) &= 
\sum_{\boldsymbol\sigma \in \mathbf{R}}  \left( \omega_{\textsc{fc}}  \sum_{i=1}^{|\sigma|-1} F^\textsc{f}_{\sigma_{i}\sigma_{i+1}}(v_{\sigma_{i}\sigma_{i+1}}) + \omega_{\textsc{fd}} \ ( t_{\sigma_{|\sigma|}} - t_{\sigma_1} ) \right)
  \label{PRPObj2}
\end{align}

\section{The proposed speed and departure time optimization algorithm}
\label{Algorithm}

This section deals with the optimization of speeds and departure times for a fixed route $\sigma$.
To simplify the exposition, we will omit $\sigma$ in the notations, and thus assume that customers are indexed by their order of appearance in the route.

The fuel consumption per distance unit $F^\textsc{f}_{i,i+1}(v_{i,i+1})$ is a convex function. The speed value $v^*_\textsc{f}$ that minimizes fuel costs is given in Equation (\ref{minF}). Similarly, for any arc $(i,{i+1})$, assuming that there is no waiting time in the route after ${i}$, the speed value $v^*_\textsc{fd}$ that minimizes fuel plus driver costs is expressed in Equation~(\ref{minFD}).

\begin{equation}
\frac{d F^\textsc{f}_{i,i+1}}{d v_{i,i+1}} (v^*_\textsc{f}) = 0  
\Leftrightarrow  v^*_\textsc{f} = \left( \frac{w_1}{2 w_4} \right)^{1/3} \label{minF}
\end{equation}
\vspace{.55cm}
\begin{equation}
v^*_\textsc{fd} = \left( \frac{\frac{\omega_\textsc{dc}}{\omega_\textsc{fc}}  + w_1}{2 w_4} \right)^{1/3} \label{minFD}
\end{equation}

\vspace{.35cm}
For a fixed route, the speed and departure time optimization problem consists of finding the departure time from the depot and the optimal speeds for each arc while respecting customers' time windows. To solve this problem, we propose an optimal recursive algorithm that extends those presented in \cite{DemirBL12,Hvattumetal2013} and \cite{KramerSVC14}. It addresses, in quadratic time, a special case of the time-dependent SOP of \cite{Franceschettietal2013}.

The algorithm relies on a general divide-and-conquer strategy, which iteratively solves a relaxed SOP obtained by ignoring time windows at intermediate destinations. If the resulting solution satisfies all constraints, then it is returned. Otherwise the customer $p$ with maximum time-window violation is identified and its arrival time is set to its closest feasible value. Fixing this decision variable creates two sub-problems which are recursively solved (\mbox{Alg. \ref{ArrivalTimes}}, \mbox{Lines 20-21}). The novelty of this algorithm is the way it manages departure or arrival-time fixing within subproblems to converge towards optimal departure and speed decisions.

\begin{algorithm}[htb]
\caption{Speed and departure-time optimization algorithm (SDTOA)}
\begin{algorithmic}[1]
  \State Procedure $SDTOA(s,e)$
  \State $p \leftarrow violation \leftarrow maxViolation \leftarrow 0$ \vspace{.05cm}
  \State $D \leftarrow \sum_{i=s}^{e-1}d_{{i},{i+1}}$ \vspace{.05cm}
  \State $T \leftarrow \sum_{i=s}^{e-1}\tau_{{i}}$ \vspace{.05cm}
  \If {$s = 1$ \textbf{and} $e = n_\sigma$}\label{SOA:computeSpeedsBegin}
    \State $t_{{1}} = a_{{1}}$
  \EndIf
  \If {$e = n_\sigma$}
    \State $t_{{e}} = \min\{\max\{a_{{e}},t_{{s}} + D/v^{*}_\textsc{fd} + T\},b_{{e}}\}$
  \EndIf
  \If {$s = 1$}
    \State $t_{{s}} = \min\{\max\{a_{{s}},t_{{e}} - D/v^{*}_\textsc{fd} - T\},b_{{s}}\}$
  \EndIf
  \State $v_\textsc{ref} \leftarrow D/(t_{{e}} - t_{{s}} - T)$\label{SOA:refSpeed}
  \For {$i = s+1 \dots e$}
    \State $t_{{i}} = t_{{i-1}} + \tau_{{i-1}} + d_{{i-1},{i}}/v_\textsc{ref}$
    \State $violation = \max\{0, t_{{i}} - b_{{i}}, a_{{i}} - t_{{i}}\}$\label{SOA:maxViolation0}
    \If {$violation > maxViolation$}\label{SOA:maxViolation1}
      \State $maxViolation = violation$\label{SOA:maxViolation2}
      \State $p = i$\label{SOA:maxViolation3}
    \EndIf
  \EndFor  
  \If {$maxViolation > 0$}
    \State $t_{{p}} = \min\{\max\{a_{{p}},t_{{p}} \},b_{{p}}\}$\label{SOA:updateArrivalTime}
    \State $SDTOA(s,p)$ \label{SOA:SOAsubProb1}
    \State $SDTOA(p,e)$\label{SOA:SOAsubProb2}
  \EndIf
  \If {$s = 1$ \textbf{and} $e = n_\sigma$} 
  \For {$i = 2 \dots n_\sigma$}
    \State \mbox{$v_{{i-1},{i}} {\small{=}} max\{d_{{i-1},{i}}/(t_{{i}} {\small{-}} t_{{i-1}} {\small{-}} \tau_{{i-1}})), v_\textsc{f}^{*}\}$} \label{SOA:computeSpeeds}
  \EndFor
\EndIf \label{SOA:computeSpeedsEnd}
\end{algorithmic}
\label{ArrivalTimes}
\end{algorithm}

For a route with $n_\sigma$ nodes (including the departure and return to the depot), Alg.~\ref{ArrivalTimes}  is applied by setting the \emph{start} $s$ to $1$ and the \emph{end} $e$~to~$n_\sigma$. The departure time is first set to the earliest possible value $t_{{1}} = a_{{1}}$ (\mbox{Alg. \ref{ArrivalTimes}}, \mbox{Line 6}). This decision will be revised later~on. 
The arrival times at each customers are then derived as follows. The arrival time at the last customer when traveling at speed $v^{*}_\textsc{fd}$ is determined and, in case of violation, updated to its closest time-window bound (\mbox{Alg. \ref{ArrivalTimes}}, \mbox{Line 8}). This leads to a reference speed $v_\textsc{ref}$ on the route  (\mbox{Alg. \ref{ArrivalTimes}}, \mbox{Line 11}) which is used to compute the arrival time at each customer as well as the maximum time-window violation (\mbox{Alg. \ref{ArrivalTimes}}, \mbox{Lines 12-17}). 

In case of violation, two subproblems are recursively solved. Any subproblem starting at the depot is now solved without fixing the departure time. Indeed, the arrival time to the last customer of this sub-problem is already fixed, such that it is possible to evaluate the reference speed ``backwards'', deriving the best departure time at the depot (\mbox{Alg. \ref{ArrivalTimes}}, \mbox{Line 10}), and the customer arrival times. The other sub-problems are similarly solved. The recursion is repeated until all constraints are satisfied. Finally, when arrival times are known for all customers, the associated speeds are revised in such a way that any speed below $v_\textsc{f}^{*}$ is replaced by $v_\textsc{f}^{*}$ and a waiting time (\mbox{Alg. \ref{ArrivalTimes}}, \mbox{Lines 22-24}).

\begin{figure}[!ht]
 \vspace{.5cm}
 \begin{center}
 \hspace*{-0.1cm}
 \includegraphics[width=16.35cm,height=7cm,]{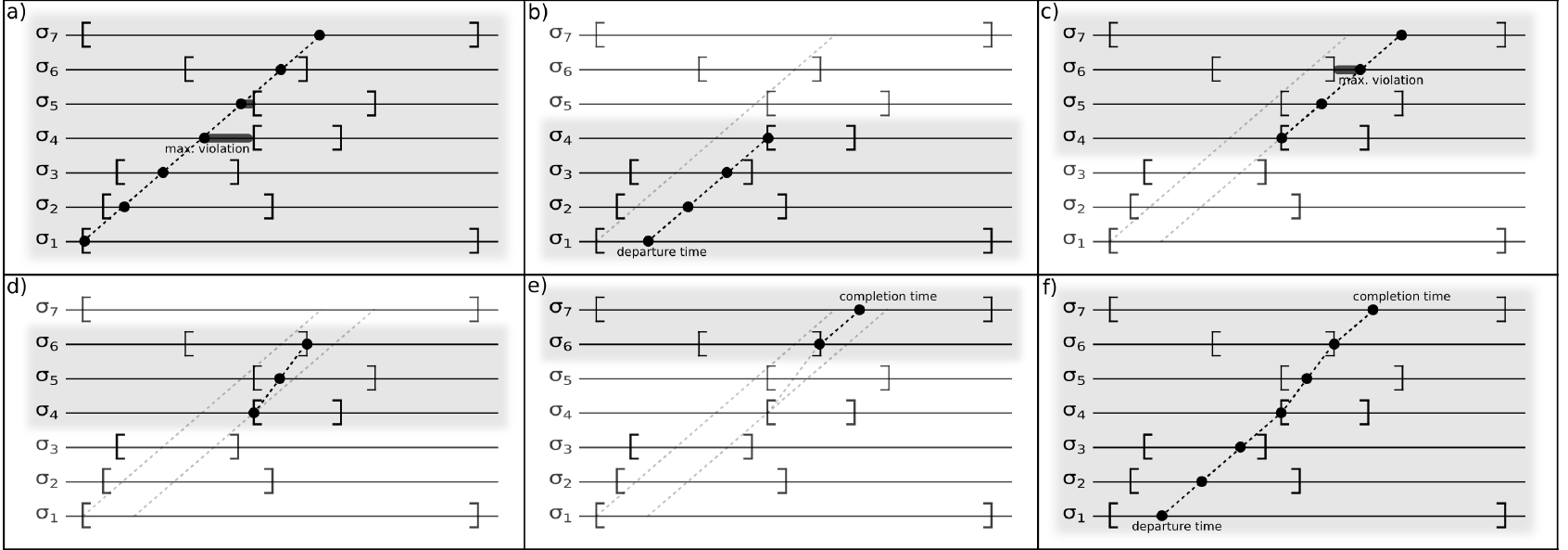}
 \caption{Computing departure time and arrival times with SDTOA}
\label{sdtoaFig}
 \end{center}
\end{figure}

Figure \ref{sdtoaFig} shows an execution example of the presented algorithm in a route involving seven nodes. The horizontal lines represent the customers and the brackets their corresponding time windows. Bullet points indicate the arrival times. The best departure and arrival times of this example are depicted in Figure \ref{sdtoaFig}.f.

\section{Proof of Optimality}
\label{ProofSDTOA}

At each iteration, for a range of indices $(s,\dots,e)$, the algorithm generates a solution with uniform speed starting from the earliest feasible time $a_{s}$ (Lines 6 and 10 of Algorithm~1). The symmetric case (Line 8, where the arrival time is fixed) reduces to a fixed starting time with a change of variable. Then, the algorithm evaluates the maximum quantity of infeasibility, either earliness or tardiness, when using the speed $v_\textsc{ref}$ on the route.
If no infeasibility is found, then the schedule is optimal. Otherwise, there exists one service time $t_K$, $K \in \{s+1,\dots,e-1\}$, with maximum constraint violation. This service time is fixed to its closest bound: $t_{K} = a_{K}$ in case of earliness, or $t_{K} = b_{K}$ in case of tardiness. Proving that at least one optimal solution exists, with this fixed variable, at each iteration would lead to the optimality of the algorithm. We will prove Theorem 1:

\vspace{.2cm}
\begin{thm}
\label{theor1}
{\normalfont \textbf{A)}} If the selected variable $K$ corresponds to a late service, then there exists no optimal solution of the subproblem $\mathbf{t^*}$ such that $t^*_{K} < b_{K}$.

\vspace{.2cm}
\hspace*{1.949cm} {\normalfont \textbf{B)}}  If the selected variable $K$ corresponds to an early service, then there exists no optimal solution of the subproblem $\mathbf{t^*}$ such that $t^*_{K} > a_{K}$.
\end{thm}

\vspace{.2cm}
\begin{myproof}
The speed optimization subproblem is reformulated in Equations~(\ref{sop1}--\ref{sop4}).
The function $c(x) = \omega_{\textsc{fc}} (w_1 x +  w_4/x^2) +  \omega_{\textsc{dc}} x$ is convex and smooth. Other constant members of the objective have been eliminated.
\end{myproof}
\begin{align}
\min\limits_{t_s,\dots,t_e}  \hspace{0.3cm} & \sum\limits_{i=s+1}^{e}  d_{i-1,i} \ \hat{c} \left( \frac{t_{i} - t_{i-1}}{d_{i-1,i}} \right) & \label{sop1} \\
\text{s.t.} \hspace{0.4cm} &  a_i \leq t_i \leq b_i  & s \leq i \leq e  \label{sop2} \\
\text{with}  \hspace{0.4cm} & \hat{c}(x) =
\begin{cases}
+\infty  & \text{for} \  x \in [0, \frac{1}{v_{max}}[ \\
c(x)  & \text{for} \ x \in  [\frac{1}{v_{max}},\frac{1}{v^*_\textsc{f}}[  \\
c(\frac{1}{v^*_\textsc{f}}) +  \omega_{\textsc{dc}} \left(x - \frac{1}{v^*_\textsc{f}} \right) &  \text{for} \ x \in [ \frac{1}{v^*_\textsc{f}},+\infty[ 
\end{cases} \label{sop4}
\end{align}

\vspace{.3cm}
Function $\hat{c}(\cdot)$ returns the best per-mile cost, including possible waiting decisions. This function is also convex and smooth, and attains its minimum for $x = 1/v^*_\textsc{fd}$.
The necessary and sufficient Karush-Kuhn-Tucker (KKT) optimality conditions of this problem can be stated as follows.
Associate the Lagrangian multipliers $(\mu^*_s, \dots,\mu^*_e)$ to the constraints $a_i \leq t_i$, and  $(\lambda^*_s, \dots,\lambda^*_e)$  to the constraints $t_i \leq  b_i$. A solution $\mathbf{t^*}$ of Equations (\ref{sop1}--\ref{sop2}) is optimal if and only if there exists $(\boldsymbol\mu^*,\boldsymbol\lambda^*)$ such that:
\begin{equation}
\label{opt}
\left\{
\begin{aligned}
& a_i \leq t_i \leq b_i  & \hspace{1cm} i \in \{s, \dots, e\}   \\
& -\hat{c}' \left(  \frac{t_{s+1} - t_{s}}{d_{s,s+1}} \right) + \lambda_s - \mu_s = 0 \\
& \hat{c}' \left(  \frac{t_i - t_{i-1}}{d_{i-1,i}} \right)  -  \hat{c}' \left(  \frac{t_{i+1} - t_{i}}{d_{i,i+1}} \right) + \lambda_i - \mu_i = 0 & \hspace{1cm} i \in \{s+1, \dots, e-1\} \\
&  \hat{c}' \left(  \frac{t_e - t_{e-1}}{d_{e-1,e}} \right)  + \lambda_e - \mu_e = 0 \\
& \lambda_i (t_i - b_i) = 0 \ ; \ \lambda_i \geq 0  & \hspace{1cm} i \in \{s, \dots, e\} \\
& \mu_i (t_i - a_i) = 0 \ ; \ \mu_i \geq 0 & \hspace{1cm} i \in \{s, \dots, e\}
\end{aligned}
\right.
\end{equation}

These optimality conditions can be reformulated as in Equations (\ref{comp1}-\ref{comp3}).

\begin{align}
\hspace*{2.8cm} \text{for } i = s, \hspace*{0.35cm}
&\left\{
\begin{tabular}{@{}rl}
either  $t_s \in ]a_s,b_s[$ &and  $\frac{t_{s+1} - t_{s}}{d_{s,s+1}} = \frac {1}{v^*_\textsc{fd}}$ \\
or $t_s = a_s$ \hspace*{0.42cm} &and $\frac{t_{s+1} - t_{s}}{d_{s,s+1}} \leq  \frac {1}{v^*_\textsc{fd}}$ \\
or $t_s = b_s$ \hspace*{0.42cm} &and $\frac{t_{s+1} - t_{s}}{d_{s,s+1}} \geq  \frac {1}{v^*_\textsc{fd}}$
\end{tabular} \right. \label{comp1} \\[2.0ex]
%
\text{for each } i \in \{s+1,\dots,e-1\},  \hspace*{0.35cm}
&\left\{
\begin{tabular}{@{}rl}
either  $t_i \in ]a_i,b_i[$ &and  $\frac{t_i - t_{i-1}}{d_{i-1,i}} =  \frac{t_{i+1} - t_{i}}{d_{i,i+1}}$ \\
or $t_i = a_i$  &and $\frac{t_i - t_{i-1}}{d_{i-1,i}} \geq  \frac{t_{i+1} - t_{i}}{d_{i,i+1}}$ \\
or $t_i = b_i$  &and $\frac{t_i - t_{i-1}}{d_{i-1,i}} \leq  \frac{t_{i+1} - t_{i}}{d_{i,i+1}}$
\end{tabular} \right. \label{comp2} \\[2.0ex]
\text{for } i = e,  \hspace*{0.35cm}
&\left\{
\begin{tabular}{@{}rl}
either  $t_e \in ]a_e,b_e[$ &and  $\frac{t_{e} - t_{e-1}}{d_{e-1,e}} = \frac {1}{v^*_\textsc{fd}}$ \\
or $t_e = a_e$ \hspace*{0.42cm} &and $\frac{t_{e} - t_{e-1}}{d_{e-1,e}} \geq  \frac {1}{v^*_\textsc{fd}}$ \\
or $t_e = b_e$ \hspace*{0.42cm} &and $\frac{t_{e} - t_{e-1}}{d_{e-1,e}} \leq  \frac {1}{v^*_\textsc{fd}}$
\end{tabular} \right.  \label{comp3}
\end{align}

\vspace{.25cm}
Now, suppose that the algorithm found a maximum tardiness for the index $K$, and that there exists an optimal solution $\mathbf{t^*}$ such that $t^*_{K} < b_{K}$. We will show by contradiction that such a solution cannot be optimal.

Let the index $I$ be defined as $I = \max \{ \{ i \ | \ i < K \text{ and } t^*_i = b_i \}\cup\{s\} \}$, that is, the largest index $i$ smaller than $K$ such that $t^*_i = b_i$, or $s$ if no such index exists. As~such,~either $ t^*_I = b_I$ \textbf{or} $\{I = s$ \textbf{and} $t^*_s < b_s\}$.

Similarly, let the index $J$ be defined as $J =  \min \{ \{ j \ | \ j > K \text{ and } t^*_j = b_i \}\cup\{e\} \}$.
As such, either $ t^*_J = b_J$ \textbf{or} $\{J = e$ \textbf{and} $t^*_J < b_J\}$. 

From the definition of $I$ and $J$, it follows that $t^*_i < b_i$ for $i \in \{I+1,\dots,J-1\}$, and thus, from the optimality conditions of Equation (\ref{comp2}), we obtain:

\begin{equation}
\frac{t^*_i - t^*_{i-1}}{d_{i-1,i}} \geq  \frac{t^*_{i+1} - t^*_{i}}{d_{i,i+1}} \text{ for } i \in \{I+1,\dots,J-1\} \label{statement}
\end{equation}

Now, recall that $K$ is the maximum tardiness obtained from a solution with constant speed $v_\textsc{ref}$, such that $v_\textsc{ref} \geq v^*_\textsc{fd}$, starting from $a_s$. 
It follows that,
\begin{align}
a_s + \sum_{i=s}^{K-1} \frac{d_{i,i+1}}{v_\textsc{ref}} - b_K \geq  a_s + \sum_{i=s}^{I-1} \frac{d_{i,i+1}}{v_\textsc{ref}} - b_I &\hspace*{0.5cm} \Rightarrow \sum_{i=I}^{K-1} \frac{d_{i,i+1}}{v_\textsc{ref}} \geq  b_K - b_I \label{eqq1} \\
a_s + \sum_{i=s}^{K-1} \frac{d_{i,i+1}}{v_\textsc{ref}} - b_K \geq  a_s + \sum_{i=s}^{J-1} \frac{d_{i,i+1}}{v_\textsc{ref}} - b_J &\hspace*{0.5cm}  \Rightarrow \sum_{i=K}^{J-1} \frac{d_{i,i+1}}{v_\textsc{ref}}  \leq   b_J - b_K.  \label{eqq2}
\end{align}

\noindent
$\bullet$
If $t^*_I = b_I$, then $b_{K} - b_{I} > t^*_{K}  - t^*_{I}$. Equation (\ref{eqq1}) leads to:
\begin{align}
& \sum_{i=I}^{K-1} \frac{d_{i,i+1}}{v_\textsc{ref}} \geq  b_K - b_I > t^*_{K}  - t^*_{I}.  \nonumber \\
\text{ with } &t^*_{K}  - t^*_{I} =  \sum_{i=I}^{K-1} (t^*_{i+1}  - t^*_{i}) =  \sum_{i=I}^{K-1} d_{i,i+1} \times \frac{t^*_{i+1}  - t^*_{i}}{d_{i,i+1}}  \nonumber \\
\Rightarrow \hspace*{0.3cm}  &  \sum_{i=I}^{K-1}    \frac{d_{i,i+1}}{v_\textsc{ref}} >   \sum_{i=I}^{K-1} d_{i,i+1} \times \frac{t^*_{i+1}  - t^*_{i}}{d_{i,i+1}}  \nonumber \\
\Rightarrow \hspace*{0.3cm}  & \boxed{ \exists \ i \in  \{I,\dots,K-1\} \text{ such that }   \frac{1}{v_\textsc{ref}} > \frac{t^*_{i+1}  - t^*_{i}}{d_{i,i+1}}.}
\label{myeq1}
\end{align}

\noindent
$\bullet$ Else $\{I = s$ and $t^*_s < b_s\}$. We have that:
\begin{align}
&t^*_K - t^*_s < b_K - a_s \leq  \sum_{i=s}^{K-1} \frac{d_{i,i+1}}{v_\textsc{ref}}, \text{and similarly} \nonumber \\
\Rightarrow \hspace*{0.3cm} &  \boxed{ \exists \ i \in  \{I,\dots,K-1\}  \text{ such that }   \frac{1}{v_\textsc{ref}} > \frac{t^*_{i+1}  - t^*_{i}}{d_{i,i+1}}.
}
\label{myeq2}
\end{align}

\noindent
$\bullet$ If $t^*_J = b_J$, then $b_{J} - b_{K} < t^*_{J}  - t^*_{K}$. Equation (\ref{eqq2}) leads to:
\begin{align}
& \sum_{j=K}^{J-1} \frac{d_{j,j+1}}{v_\textsc{ref}}  \leq  b_J - b_K < t^*_{J}  - t^*_{K}.   \nonumber \\
\text{ with }& t^*_{J}  - t^*_{K}  =  \sum_{j=K}^{J-1}   (t^*_{j+1}  - t^*_{j}) =  \sum_{j=K}^{J-1} d_{j,j+1} \times  \frac{t^*_{j+1}  - t^*_{j}}{d_{j,j+1}}   \nonumber \\
\Rightarrow \hspace*{0.3cm} &   \sum_{j=K}^{J-1}   \frac{d_{j,j+1} }{v_\textsc{ref}}  <  \sum_{j=K}^{J-1} d_{j,j+1} \times  \frac{t^*_{j+1}  - t^*_{j}}{d_{j,j+1}}   \nonumber \\
\Rightarrow \hspace*{0.3cm} & \boxed{\exists \ j \in \{K,\dots,J-1\} \text{ such that }    \frac{1}{v_\textsc{ref}} < \frac{t^*_{j+1}  - t^*_{j}}{d_{j,j+1}}.} \label{myeq3}
\end{align}

\noindent
$\bullet$ Else $\{J = e$ and $t^*_e < b^*_e\}$, then the condition of Equation (\ref{comp3}) states that
$ \frac {1}{v^*_\textsc{fd}} \leq \frac{t^*_{e} - t^*_{e-1}}{d_{e-1,e}}$. Since $v^*_\textsc{fd}  \leq v_\textsc{ref}$, this leads to the same statement as Equation (\ref{myeq3}).

Overall, Equations (\ref{myeq1}--\ref{myeq2}) and  (\ref{myeq3})  lead to:
\begin{equation}
\boxed{ \exists \ (i,j) \ | \ i < K \leq j,  \text{ such that }   \frac{t^*_{i+1}  - t^*_{i}}{d_{i,i+1}}  <  \frac{1}{v_\textsc{ref}}  \leq \frac{t^*_{j+1}  - t^*_{j}}{d_{j,j+1}}.}
\end{equation}

This is a direct contradiction of Equation (\ref{statement}).
Our original assumption, that there exists an optimal solution with $t^*_{K} < b_{K}$ is impossible, hence completing the proof of statement \textbf{A)}. The proof of case \textbf{B)} is analogous. $\hfill \qquad \Box$

\vspace*{-0.2cm}
\section{Computational Experiments}
\label{compExpSDTOA}

The proposed algorithm was integrated into the ILS-SP-SOA matheuristic of \cite{KramerSVC14}, 
which makes use of an adaptive speed matrix during the search to keep track of speed decisions.
The speed and departure time optimization algorithm is executed on each route associated to a local optimum of ILS. These routes are also stored in a pool and used to generate new solutions by means of integer programming over a set partitioning (SP) formulation.

We tested the method on the instances of \cite{DemirBL12} and \cite{KramerSVC14}, containing between $10$ and $200$ customers. 
The coefficients of the objective function were set to the same values as \cite{KramerSVC14}.
The algorithm was implemented in C++ and executed on an Intel i7 3.40 GHz processor with 16 GB of RAM, using CPLEX 12.4.

Table \ref{avgGaps} reports the results of these experiments. Ten independent runs were done for each instance, and each line corresponds to averaged results on a set of 20 instances. All detailed results are available at \url{http://w1.cirrelt.ca/~vidalt/en/VRP-resources.html}. The new solutions are compared to the previous best known solutions (BKS) from \cite{KramerSVC14}, which did not consider departure-time delays. The rightmost part of the table also reports the effects of departure time optimization, in percentage, on labor costs, fuel and distance (comparing our best solution to the BKS).

\begin{table}[h]
\centering
\setlength{\tabcolsep}{0.25cm}
\renewcommand{\arraystretch}{1.4}
\caption{Results for the PRP with departure time optimization}
\scalebox{.78}
{
\begin{tabular}{lcccccc@{\hspace{1.1cm}}ccc}
\hline
\textbf{} & \multicolumn{ 3}{c}{\textbf{ILS-SP-SDTOA}} & \textbf{} & \textbf{} &  & \multicolumn{ 3}{c}{\textbf{Effects(\%) on:}} \\ \cline{2-4} \cline{8-10}
\textbf{\up{Inst.}} & \textbf{Avg-10} & \textbf{T(s)} & \textbf{Best-10} & \textbf{\up{BKS}} & \textbf{\up{Gap(\%)}} &  & \textbf{Labor} & \textbf{Fuel} & \textbf{Dist.} \\ \hline
10-A & 183.14 & 0.04 & \textbf{183.06} & 185.50 & -1.44 &  & -3.17 & -0.26 & -0.16 \\ 
50-A & 601.18 & 3.16 & \textbf{600.26} & 609.71 & -1.59 &  & -2.71 & -0.75 & -0.67 \\ 
100-A & 1105.12 & 36.20 & \textbf{1102.83} & 1119.17 & -1.48 &  & -2.60 & -0.62 & -0.54 \\ 
200-A & 1951.74 & 340.71 & \textbf{1944.75} & 1964.81 & -1.04 &  & -1.94 & -0.29 & -0.27 \\ \cline{1-6} \cline{ 8-10}
10-B & 248.22 & 0.04 & \textbf{247.88} & 281.15 &  -11.69 &  & -23.24 & 0.51 & -0.92 \\ 
50-B & 771.03 & 4.76 & \textbf{770.29} & 849.96 & -9.53 &  & -20.49 & 1.42 & -0.08 \\ 
100-B & 1386.35 & 69.57 & \textbf{1384.70} & 1536.19 &  -9.87 &  & -1.94 & -0.29 & -0.27 \\ 
200-B & 2345.41 & 782.76 & \textbf{2334.08} & 2688.90 & -13.22 &  & -28.15 & 4.25 &  2.25 \\ \cline{1-6} \cline{ 8-10}
10-C & 195.21 & 0.04 & \textbf{195.08} & 222.16 & -12.29 &  & -26.78 & 0.60 & -1.57 \\ 
50-C & 672.86 & 4.77 & \textbf{671.83} & 755.44 & -11.23 &  & -23.57 & 0.86 & -1.20 \\ 
100-C & 1228.35 & 61.13 & \textbf{1226.13} & 1398.56 & -12.32 &  & -25.44 & 0.92 & -1.21 \\
200-C & 2114.28 & 482.29 & \textbf{2103.18} & 2461.05 & -14.58 &  & -28.59 & 1.23 & -1.25 \\ \cline{1-6} \cline{ 8-10}
\textbf{Avg.} & \textbf{1066.91} & \textbf{148.79} & \textbf{1063.67} & \textbf{1172.72} & \textbf{-8.36} & \textbf{} & \textbf{-15.72} & \textbf{0.63} & \textbf{-0.49} \\ \hline
\end{tabular}
\label{avgGaps}
}
\end{table}

\begin{figure}[!b]
 \begin{center}
  \includegraphics[width=0.99\textwidth]{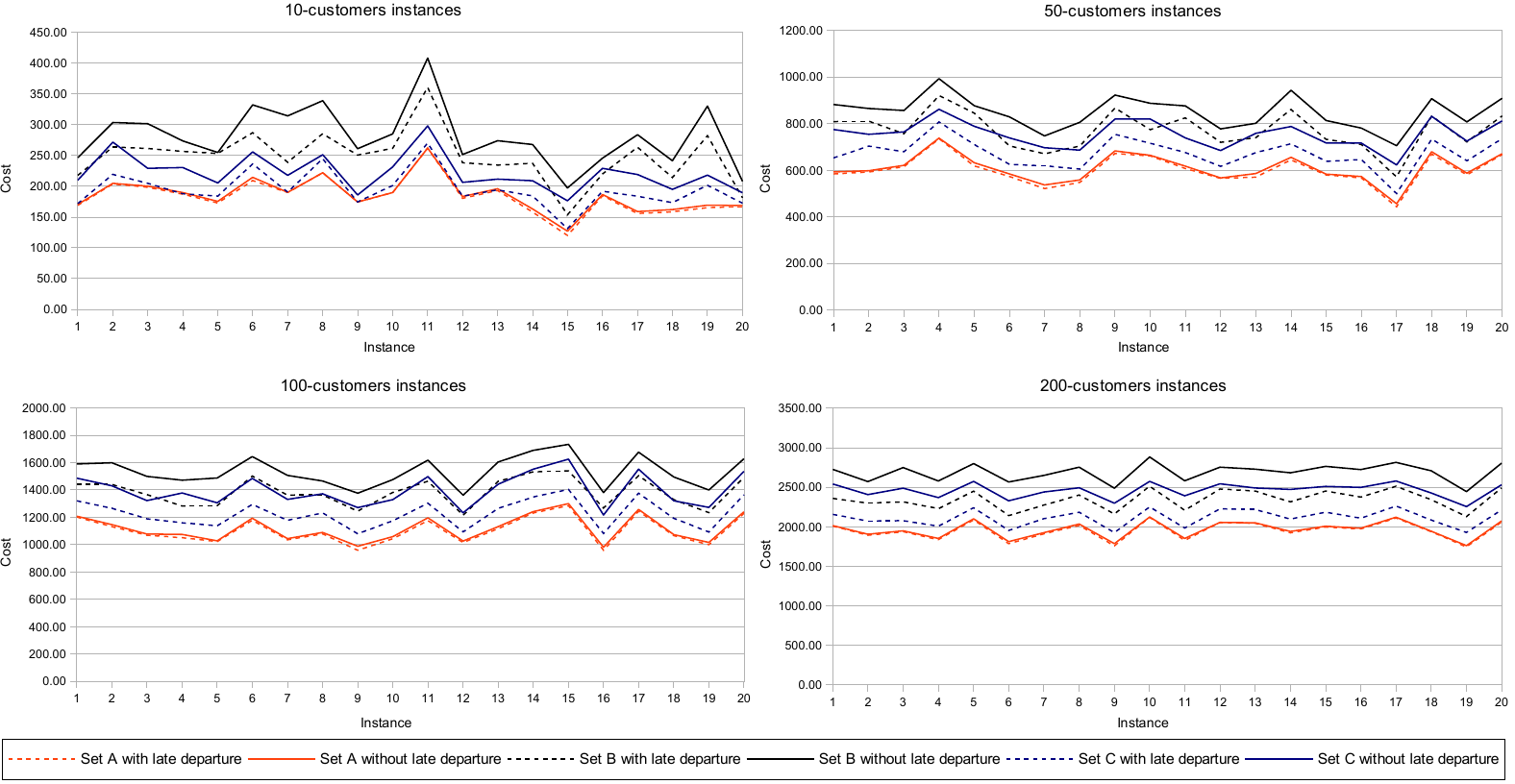}\\
  \caption{Comparison of solution costs, with or without delayed departure times.}
\label{costcomparison1}
 \end{center}  
\end{figure}

In light of these results, allowing delayed departures times can lead to significant reductions of operational costs ($-8.36\%$ on average).  Instances with tighter time-windows (sets $B$~and~$C$) tend to be more prone for solution quality improvement based on delays at the depot, as illustrated in Figure \ref{costcomparison1} for individual instances.

The main cost improvements originate, as expected, from reduced labor costs ($-15.72\%$).
Moreover, we also notice a quasi-systematic decrease of driven distance ($-0.49\%$). Delayed departure time optimization indeed leads to a larger set of good-quality candidate routes, increasing the size of the search space, resulting in further opportunities of distance minimization. 

\begin{figure}[ht]
 \begin{center}
  \includegraphics[width=0.92\textwidth]{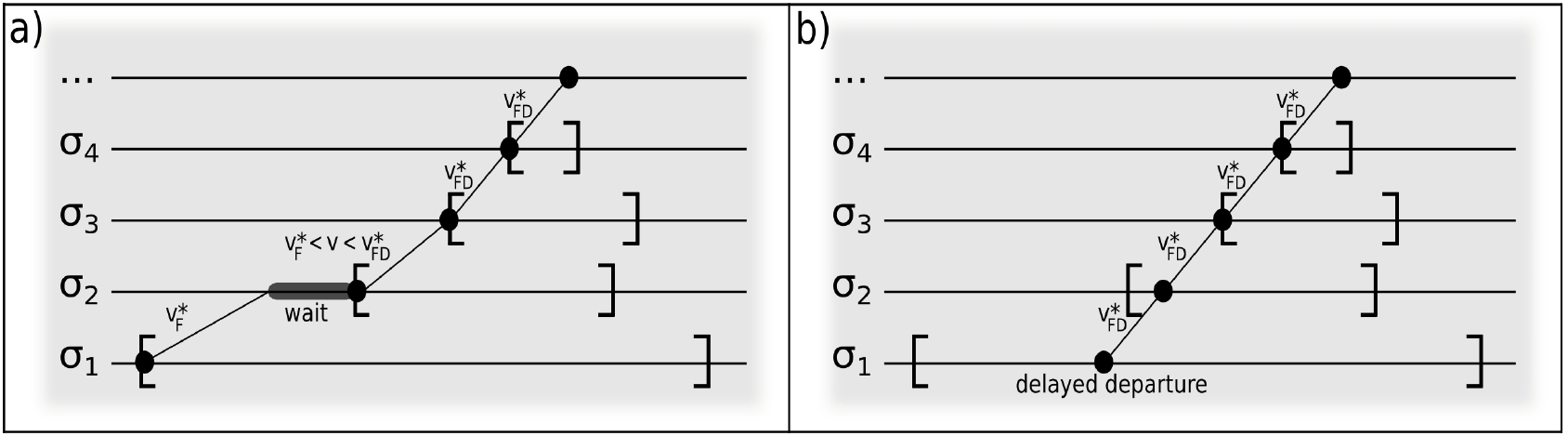}\\
  \caption{Different speed choices with or without departure time optimization.}
\label{speedschg}
 \end{center}  
\end{figure}

The impact on pollution emissions depends on the tightness of the time constraints.
For all benchmark sets of type A, with large time windows, a reduction can be observed ($-0.48\%$) as a consequence of the improvement in total distance.
For most benchmark sets of type B and C, with tight time windows, we observe a moderate but consistent increase of fuel consumption ($+1.18\%$). This is an indirect consequence of the optimality conditions of the SOP. Indeed, when confronted with an active time-window constraint of the type $a_i \leq t$, early in the route, the speed optimization algorithm with fixed departure time creates a waiting time followed by a travel at speed $v^*_\textsc{f}$ (minimal emissions since the labor costs are constant).
This is illustrated on the left of Figure~\ref{speedschg}.  

In contrast, the speed and departure time optimization algorithm will delay the departure, using the speed $v^*_\textsc{fd}$ (minimal emissions + labor costs, such that $v^*_\textsc{fd} \geq v^*_\textsc{f}$)  to arrive exactly on time to the first active constraint, as illustrated on the right part of Figure \ref{speedschg}.
This effect is particularly visible on instances with more induced waiting times and opportunities of labor cost reductions.

\section{Conclusions}
\label{Conclusions}

A new speed and departure time optimization algorithm for the PRP has been presented.
This algorithm is conceptually simple, runs in quadratic time, and is guaranteed to produce an optimal solution. It was implemented and integrated in the matheuristic of~\cite{KramerSVC14}, where departure time and speed optimizations occur for each local minimum of the iterated local search. Our experimental results with this heuristic showed that delayed departure times from the depot can lead to very significant savings: up to $8.36\%$ operational costs for the considered benchmark~sets.

Overall, integrated scheduling, speed control, and routing optimization can help to reduce costs and environmental fingerprints in a variety of other industrial domains.
The proposed methodology has contributed to address some open challenges related to combined speed and schedule optimization. Further research can now be focused on generalizing these methods to broader application classes. In particular, arc-dependent cost/speed functions are very relevant for ship operations in the presence of variable weather and sea conditions.  However, no efficient algorithm is known for~this~setting.

\section*{Acknowledgments}
\label{acknowledgments}

This research was partially supported by the Instituto Nacional de Ciência e Tecnologia (INCT) 
and by the Conselho Nacional de Desenvolvimento Científico e Tecnológico (CNPq), grants \linebreak 557128/2009-9, 471158/2012-7, and \sloppy{GDE 201222/2014-0}.

\bibliography{ref}

\end{document}